\documentclass{llncs}
\usepackage{makeidx}
\usepackage{graphics}
\usepackage{url}

\begin{document}
\frontmatter
\pagestyle{headings}
\addtocmark{Internet Adversary}
\mainmatter
\title{Notes On The Design Of An Internet Adversary}
\titlerunning{Internet Adversary}

\author{David S. H. Rosenthal\inst{1}
\and
Petros Maniatis\inst{2}
\and
Mema Roussopoulos\inst{3}
\and
TJ Giuli\inst{4}
\and
Mary Baker\inst{5}
}
\authorrunning{David Rosenthal et al.}
\tocauthor{David S. H. Rosenthal (Stanford),
Petros Maniatis (Intel Research),
Mema Roussopoulos (Harvard),
TJ Giuli (Stanford),
Mary Baker (HP Labs)}

\institute{Stanford University Libraries, Stanford, CA,
\and
Intel Research, Berkeley, CA,
\and
Computer Science, Harvard University, Cambridge, MA,
\and
Computer Science Department, Stanford University, Stanford, CA,
\and
HP Labs, Palo Alto, CA }
\maketitle

\begin{abstract}
The design of the defenses Internet systems can deploy against attack,
especially adaptive and resilient defenses, must start from a
realistic model of the threat.  This requires an assessment of the
capabilities of the adversary.  The design typically evolves through a
process of simulating both the system and the adversary.  This
requires the design and implementation of a simulated adversary based
on the capability assessment.  Consensus on the capabilities of
a suitable adversary is not evident.  Part of the recent redesign
of the protocol used by peers in the LOCKSS digital preservation
system included a conservative assessment of the adversary's
capabilities.  We present our
assessment and the implications we drew from it as a step towards
a reusable adversary specification.
\end{abstract}

\section{\label{sec:introduction}Introduction}

The LOCKSS\footnote{LOCKSS is a trademark of Stanford University.}
(Lots Of Copies Keep Stuff Safe) program has developed and deployed
test versions of a system for preserving access to academic journals
published on the Web.  The fundamental problem for any digital
preservation system is that it must be affordable for the long term.
To reduce the cost of ownership, the LOCKSS system uses generic PC
hardware, open source software, and peer-to-peer technology.  It is
packaged as a ``network appliance,'' a single-function box that can be
connected to the Internet, configured and left alone to do its job
with minimal monitoring or administration.  The system has been under
test at about 50 libraries worldwide since 2000.

Like other Internet hosts, these appliances are continually subject to
attack.  Although measures~\cite{Rosenthal2003b} have been taken to
render the operating system platform resistant to attack, its
compromise must be anticipated.  The appliances cooperate with each
other to detect and repair damage in a peer-to-peer network.  The
first version~\cite{Rosenthal2000} of this protocol turned out to be
vulnerable to various attacks.  We recently redesigned the
protocol~\cite{Maniatis2003lockssSOSP} to make it more resistant to attack.

The redesign needed as input an assessment of the capabilities and
strategies of the potential adversaries, but we were unable to find
this information off-the-shelf.  We present our assessment, and the
implications we drew from it, as a contribution to an eventual
reusable adversary specification.

\section{\label{sec:adversaryAssessment}Adversary Assessment}

Military intelligence seeks to develop so-called ``appreciations'' of a
potential adversary's ``capabilities'' (what the adversary \emph{could}
do) and ``intentions'' (what the adversary is \emph{expected} to attempt
with the capabilities) as a basis for
planning~\cite{Army1985}.  Similarly, plans and techniques for defending
distributed systems exposed to the Internet need an appreciation of the
capabilities and intentions of the adversary they may encounter when
deployed.

Our assessment identified the following probable adversary capabilities:
\begin{itemize}
\item Unlimited Power
\item Unlimited Identities
\item Conspiracy
\item Eavesdropping and Spoofing
\item Exploiting Common Vulnerabilities
\item Uncovering Secrets
\end{itemize}

\subsection{\label{sec:unlimitedPower}Unlimited Power}

Techniques~\cite{Staniford2002} have been described by which a worm
could compromise a large proportion of vulnerable Internet hosts
in a short time.
In practice,
even much less sophisticated techniques~\cite{SlammerAnalysis} have proven
capable of compromising large numbers of hosts quickly,
despite widespread knowledge of both the vulnerabilities themselves
and their cures for
six months prior to the attack~\cite{SlammerVulnerability}.
Further:
\begin{itemize}
\item Experience with Code Red~\cite{CodeRedAnalysis} shows that at least 1/3
of the compromised hosts remain compromised a month after the start of the
attack.
Two years after the attack a pool of 20,000 infected hosts was still
available~\cite{Krebs2003}.
\item Experience with Slapper~\cite{Rescorla2003} shows that 1/3 of
vulnerable hosts were still vulnerable 3 months after the vulnerability
was announced and 1 month after the start of the attack.
\item Experience with a BIND vulnerability~\cite{BindVulnerability}
shows that a significant proportion of professionally maintained systems
are still vulnerable two months after the vulnerability was made public.
\item Advertisements are rumored to be appearing that invite spam senders
to rent access to a network of 450K compromised hosts they
can use to disguise the origin of e-mails.
\end{itemize}

So far, these networks of compromised hosts have been used to mount
crude but effective~\cite{MiMailAttack} network-level denial of
service attacks.
However,
it would be a simple matter for the payload of such a worm to be
an application-level attack targeted at a particular victim system.
If the worm were based on a vulnerability
as widespread (350K+ hosts) as the ones Code Red~\cite{CodeRedAnalysis}
or Blaster (385K+ hosts)~\cite{Krebs2003} exploited,
the attacker could expect on the order of 10K machine-years of
computation to be available for the attack on the victim system
(30\% of systems compromised for 1 month, 10\% for 3 months).
This is,
for example,
about 35 times the effort used to win the RSA DES Challenge III in
1999~\cite{DESCrack}.

There is a practical difficulty for the adversary hoping to use these
pools of compromised hosts as a resource for attacking a given system.
Many other adversaries with other targets are in competition for the
resource,  which is not infinite although it may be large.  This
difficulty, however, is not a comfort to the designer of system defenses,
whose worst-case analysis must assume that all available resources may
be used for a single-minded attack against his system.

\subsection{\label{sec:unlimitedIdentities}Unlimited Identities}

Given the relative ease by which an adversary can compromise and
control a large number of hosts across the Internet, we must assume
that the adversary can pose as an unlimited number of identities,
e.g., IP addresses. The adversary can either directly use the
compromised host's IP address or make the compromised host spoof other
IP addresses on the same subnet.  Even if ingress filtering
\cite{ingress} were turned on in all routers across the Internet, the
cost for a host to spoof an IP address on the same subnet is
negligible.

There is a practical difficulty for the adversary in that he can only
steal identities on subnets in which he maintains a presence, either
legitimately or through compromise.  This difficulty is not a comfort to
the designer of system defenses who must assume that the adversary can
have a presence in thousands of subnets spread across the Internet.

The assessment above is not unique to IP addresses.  Email addresses,
identity certificates, DNS domains are just as easy for an adversary to
hoard or spoof or both.  Techniques for making this more difficult or
time-consuming for an adversary include client puzzles and reverse
Turing tests~\cite{captcha}, but the adversaries are adapting to them.
For example,
it is now rumored that reverse Turing tests can be forwarded to a
service run by porn sites,
which exploit their customers to solve them and return their
responses.

\subsection{\label{sec:conspiracy}Conspiracy}

The Fizzer worm uses IRC~\cite{FizzerComms} to communicate with a central
control site.
It would be possible for a worm to use peer-to-peer communication techniques
instead,
avoiding the difficulties the Fizzer worm suffered
when its IRC channel was subverted by its enemies~\cite{FizzerCounter}.

It has to be assumed,
therefore,
that all the adversary's identities mask a single distributed adversary
with instantaneous self-awareness.  Any state, such as messages
sent, received, or observed by one identity acting on behalf of
the adversary is immediately made available to all other identities.

In addition,
it must be assumed that some apparently benign identities are
conspiring with the adversary.
Anything known to these ``spies,''
including supposed secrets such as session keys,
is known to the adversary.

It is practically difficult for the adversary 
to distribute information rapidly and completely among the
components of a distributed system with as many nodes as there are
compromised hosts.  This difficulty is not a comfort for the designer of
system defenses, who must assume that the adversary can succeed in
getting the critical information to the nodes that need it.

\subsection{\label{sec:eavesdropping}Eavesdropping and Spoofing}

A single compromised host on a subnet can eavesdrop on traffic to
and from all hosts on the same subnet.  It can also send spoofed
messages on behalf of the co-located hosts,
as well as send messages with spoofed source addresses from
anywhere in the Internet to
co-located hosts.  By doing so it can often abuse trust relationships
mediated by IP addresses.  This behavior is very difficult to detect
and prevent when compromised hosts are not regularly monitored and
maintained.

\subsection{\label{sec:commonVulnerabilities}Common Vulnerabilities}

Even if the design of the system's defenses is perfect, the designer
cannot assume that their implementation is as perfect.  It is likely
that, at some point, an exploitable implementation vulnerability will be
discovered.  A well-designed flash worm exploiting it can compromise
the vast majority of the vulnerable hosts in a very short time.

In different contexts including traditional Byzantine Fault
Tolerance~\cite{Castro1999}, Distributed Hash Tables~\cite{Castro2002}
and sampled voting~\cite{Maniatis2003lockssSOSP} it has been shown that
systems with more than about 1/3 faulty or malign peers cannot survive
for long.  Given this, even in fault-tolerant systems, peers need to be
assigned at random one of at least four independent implementations if
the system is to survive the discovery of an implementation
vulnerability.  Rodrigues et al.~\cite{Rodrigues2001} describe a
framework within which independent implementations can be accommodated
in a fault-tolerant system.

It is important to note that even a perfectly designed and implemented
system cannot avoid vulnerabilities brought about by human operators who
are coerced to misbehave.  An invulnerable computer system, though
unimaginably hard to build, is certainly easier to imagine than an
incorruptible human.

\subsection{\label{sec:uncoveringSecrets}Uncovering Secrets}

Most systems rely on secret-based encryption systems to preserve
system integrity.  The assumption is that the adversary does not
know and cannot in a timely fashion obtain any of the secrets.

This is not a robust assumption.  A recent survey~\cite{BallPointPasswords}
purported to show that the vast majority of commuters at
a London station would reveal their passwords if offered a ball-point
pen.  The adversary may conspire with an insider, he may be the beneficiary
of lax security by insiders such as poor password choice~\cite{Klein1990},
he may steal authentication tokens,  and, given the resources we assume, he may
even use brute-force techniques to break the encryption.

System designers should not treat encryption as a
panacea~\cite{RosenthalSeiden}.  An individual analysis is needed of the
consequences of compromise of each key in the system,  if only to assess
the precautions appropriate for its protection.

\section{\label{sec:intentions}Intentions}

We have presented an assessment of some of the putative adversary's
capabilities.  We must now assess his possible intentions.  What
might the adversary be intending to achieve by exploiting these
capabilities?

Our initial attempt classifies possible adversary intentions into
five classes: Stealth, Nuisance, Attrition, Thief, and Spy.

\subsection{\label{sec:Stealth}Stealth}

The \emph{Stealth} adversary's goal is to damage the system by affecting
its state.  A necessary sub-goal is to avoid detection before the damage
is complete, for example to dodge an intrusion detection system.

\subsection{\label{sec:nuisance}Nuisance}

The \emph{Nuisance} adversary's goal is to discredit the system by
continually raising intrusion alarms.  There is no intention to
cause any actual damage to the system or prevent it from functioning.
An attack from the Nuisance adversary might,  for example,  be
intended to get the victim's system administrators to disable or
ignore the intrusion alarms as a prelude to other forms of attack.

\subsection{\label{sec:Attrition}Attrition}

The \emph{Attrition} adversary's goal is to prevent the system from functioning
for long enough to inflict damage on the organization it supports.
Some forms of the adversary are referred to as ``Denial of Service,''
but this has come to mean a technique rather than a goal.

The Blaster worm was an Attrition attack, attempting to mount a flooding
attack on a Microsoft website from its 385K infected hosts.  The MiMail
virus is an Attrition attack against a set of anti-spam
services~\cite{MiMailAttack}.

\subsection{\label{sec:thief}Thief}

The goal of the \emph{Thief} adversary is to steal services provided by
the system (possibly over long time periods) or steal valuable
information protected by the system.  The Thief is different from the
Stealth adversary in that he does not necessarily want to alter the
state of the system, nor does he want to bring the system down or
subvert it.  The Thief of services wants unauthorized access to
resources for as long as possible without being detected.  The Thief of
information hopes that his intrusion remains undetected for as long as
possible.

The Sobig series of viruses~\cite{Sullivan2003} is believed to be a
Thief who steals services from victim machines by using them as a
spam-sending network.  It is also thought to be used to mount
Attrition attacks on anti-spam services~\cite{SobigSpam}.

\subsection{\label{sec:spy}Spy}

The \emph{Spy} adversary's goal is to observe as much about the system
as possible: who participates, where users are located, and what
transactions take place.  The Spy could be a powerful corporation
wanting to harass or prosecute users.  The Spy could also be a
government collecting information on the on-line activities of its
citizens.

\section{\label{sec:rulesOfThumb}Rules Of Thumb}

We summarize these assessments with some conservative ``rules of thumb.''
The assumptions underlying them are a worm infecting three times
as many hosts as Code Red,
with the bulk of the infection lasting four days,
and 10\% still infected after three months.
The adversary can:

\begin{itemize}

\item exert bursts of computational effort lasting 100 hours and
using 1,000,000 hosts,
\item sustain computational effort over 100 days using 100,000 hosts,
\item masquerade behind 1,000,000 IP addresses,
\item eavesdrop on and spoof traffic from 10\% of the hosts in the
victim system for 100 days.
\item break 100 well-chosen DES keys.
\end{itemize}

\section{\label{sec:implications}Implications}

Our adversary is very powerful, posing a number of important
implications.  First, it is economically infeasible to test,
or even simulate,
attacks of this scale.
Assurance that a system does not fail under expected attacks is not likely to
be available or credible.
Design should focus on:
\begin{itemize}
\item Graceful, or at least survivable, failure.
\item Assisting diagnosis, perhaps by using bimodal
behaviors~\cite{Birman1999} to raise alarms.
\item Assisting recovery.
\end{itemize}

Second, the adversary can mount a full-scale attack with no warning.
Rate-limiting techniques~\cite{Rosenthal2000,Forrest2000,Williamson2002}
are important in slowing the rate of failure enough to allow for
human intervention before failure is total.

Third, 
the adversary can appear as huge numbers of new peers or clients.
Limiting the rate at which the system accepts new peers or clients
using techniques such as ``newcomer pays''~\cite{Friedman2001} may
help slow the failure.

\section{\label{sec:relatedWork}Related Work}

Researchers in many different fields have tackled the task of
characterizing malicious adversaries.  In this section, we outline only
a few of the approaches we have identified in the literature.

\begin{itemize}

\item \emph{Cryptography} typically uses game-theoretic analyses to
construct sets of ``games'' resulting in the adversary behavior observed
by benign protocol participants, and investigate whether those sets
contain games with malign participants.

\item \emph{Protocol design} typically uses exhaustive search of the
transitive closure of the state space of the protocol without explicitly
modeling an adversary's capabilities or intentions.  Finite-state
analysis takes the same approach in an automated fashion, with some
notable successes (see, for example, an automated analysis of
authentication protocols~\cite{Mitchell1997}).

\item \emph{Distributed systems theory} typically works backwards from a
bad state of the system (e.g., a state in which an exploit has been used
to damage the system) to identify the sequence of events that must have
happened to arrive at that state.  The system has to be specified in a
suitable formalism (e.g., Lamport's TLA+~\cite{Lamport2002} or Lynch and
Tuttle's Input/Output Automata~\cite{Lynch1989,Lynch1996}), but in some
cases it is possible to conduct an invariant analysis without a full
system specification.

\item \emph{Fault tolerance} typically places broad limits on the
adversary (e.g., ``no more than 1/3 of the nodes can be malign'' in the
case of byzantine fault tolerance~\cite{Lamport1982}).  In other cases,
nodes with similar failure modes can be grouped together into distinct
equivalence classes with respect to failures (e.g., in Malkhi and
Reiter's work on quorum systems~\cite{Malkhi1998}).  These can be
loosely considered an adversary model.

\end{itemize}


Previous work on defending systems against attack classifies adversaries
as either ``computationally bounded or unbounded'' and considers the
time interval over which the adversary collects or modifies state
\cite{Dingledine2000b}.  Although the pool of vulnerable machines on
which an adversary can draw is in fact limited, it is large enough and
the repair rates low enough that the adversary may be considered
effectively unbounded in effort and time.

RFC3607~\cite{rfc3607} describes how a worm payload can be used for
cryptanalysis,
and identifies the first such payload observed in the wild.



\section{\label{sec:conclusion}Conclusion}

We have presented what we believe is a conservative assessment of
the putative adversary the designers of defenses for an Internet
system must take into account.
This adversary is based on reasonable extrapolations from
the observed behavior of worms exploiting vulnerabilities in
applications and systems that are widely deployed on the Internet,
and on the assumption that the payload of such worms might be
targeted at the system under consideration.  We believe that
discussion of this and alternative adversary assessments
leading to some consensus as a basis for future designs would
be valuable.

Our adversary is powerful enough to pose design, implementation
and testing problems well beyond those current technology can solve.
It appears that designing systems to survive attacks of this
magnitude unimpaired is unlikely to succeed.
Further,
even if the design appeared to succeed,
testing implementations to assure that success was manifest in practice
is unlikely to be affordable.  A more reasonable goal may be to
slow and delay the process of failure under attack to allow for
human intervention.

\section{\label{sec:acknowledgements}Acknowledgments}

This material is based upon work supported
by the National Science Foundation under Grant No.\ 9907296, however any
opinions, findings, and conclusions or recommendations expressed in this
material are those of the authors and do not necessarily reflect the
views of the National Science Foundation. 

The LOCKSS program is grateful for support from the National Science
Foundation,  the Andrew W. Mellon Foundation, Sun Microsystems
Laboratories, and Stanford Libraries.

Vicky Reich has made the LOCKSS program possible.

\bibliographystyle{plain}
\bibliography{../common/bibliography}


\end{document}